\documentstyle[psfrag,aps,preprint,epsfig]{revtex}

\newcommand{\Figref}[1]{Fig.~\ref{#1}}

\begin{document}

\tighten

\preprint{TU-616}
\title{Splitting Triplet and Doublet in Extra Dimensions}
\author{Mitsuru Kakizaki\footnote{e-mail: kakizaki@tuhep.phys.tohoku.ac.jp}
 and Masahiro Yamaguchi\footnote{e-mail: yama@tuhep.phys.tohoku.ac.jp}}
\address{Department of Physics, Tohoku University,
Sendai 980-8578, Japan}
\date{April 2001}
\maketitle
\begin{abstract}
A novel mechanism to realize the triplet-doublet splitting in 
supersymmetric SU(5) grand unified theories is proposed in the
framework of higher dimensional theories where chiral multiplets are
localized due to kink configuration of a SU(5) singlet. An adjoint 
Higgs field which spontaneously breaks the SU(5) gauge symmetry is assumed 
to be involved with the localization process,  splitting the wave functions 
of the color-triplet Higgs from its doublet counterpart.  The resulting 
effective four-dimensional theory does not possess manifest SU(5) 
invariance. By adjusting couplings, the doublet mass can be exponentially 
suppressed.  We also show that dimension 5 proton decay from triplet 
Higgs exchange can be suppressed to a negligible level.
\end{abstract} 

\clearpage

Supersymmetric (SUSY) grand unified theory (GUT) \cite{SUSYGUT} is a very
interesting framework for unified description of particles and gauge
forces. The gauge hierarchy problem inherent to GUT is solved by
introduction of SUSY. The gauge interaction in the standard model 
follows a unified description at very high energy scale. It is very suggestive
that 
experimental data support this view \cite{gauge-coupling}.

A simple class of SUSY GUTs, however, suffers from two severe problems.
The first one is the triplet-doublet splitting problem (in $SU(5)$
terminology). A Higgs doublet in the standard model is associated with
its color-triplet counterpart to make a complete 5 (or $5^*$)
representation in $SU(5)$.  (A similar argument applies for a larger
GUT group.) The Higgs doublet mass should be around the weak scale,
while the triplet should be superheavy. This splitting is quite
non-trivial. In the minimal SUSY $SU(5)$, it is achieved by fine
tuning parameters in the superpotential of  order $10^{14}$.  This
is possible, but obviously unsatisfactory.  The second problem is
about proton decay. In SUSY GUTs, exchange of the color-triplet Higgs
superfields generically induces dimension-five proton-decay operators 
\cite{dim5protondecay},
which will predict too short life time of the proton, inconsistent
with experimental results \cite{Goto:1999qg}.

One of the solutions for the first problem is the so-called missing
doublet mechanism based on group theoretical grounds. Examples are
given in $SU(5)$ GUT \cite{Masiero:1982fe}
and in a product GUT such as $SU(5) \times
U(3)_H$ \cite{Hotta:1996cd}.  
The second problem of the proton decay is  solved in
the product GUT  and also in an axionic extension of the
missing doublet mechanism \cite{Hisano:1995fn}.

In this paper, we would like to explore an alternative approach to the
above two problems, which is inherently higher dimensional.  The point is
that the effective four-dimensional theory obtained after dimensional
reduction need not respect $SU(5)$ symmetry if $SU(5)$ gauge symmetry
is {\em non-trivially} broken in the extra dimension(s).  To realize
this point, we utilize the localization mechanism of chiral multiplets
in five dimensions. 
Here we shall
consider the situation that, in addition to a GUT singlet which has 
a non-trivial field configuration
along the compactified  5th dimension, an
 adjoint Higgs field  responsible for
the GUT gauge symmetry breaking couples to the Higgs multiplets and its 
conjugate fields in five dimensions. 
Then the wave function of the triplet and that
of the doublet are localized at different places in the extra dimension 
and are not related with each other  by the $SU(5)$ 
transformation.  We
will show that the superpotential of the resulting effective
four-dimensional theory does not possess manifest $SU(5)$ invariance,
while the gauge coupling unification is unchanged.  Note that recently
there have been  very interesting proposals by 
\cite{Kawamura:2000nj,Altarelli:2001qj,Kobakhidze:2001yk,Hall:2001pg}.  
There the
parity assignment in five dimensions does not commute with the
$SU(5)$ symmetry and thus the resulting four-dimensional theory
does not have apparent $SU(5)$ invariance.

Using the mechanism outlined above, we shall try to explain the
triplet-doublet splitting of the Higgs multiplet. 
We will also show that the dimension five operators inducing the proton
decay can become negligibly small.

We first review the localization of chiral multiplets in five dimensional 
theory. 
The localization of fermion wave functions under solitonic backgrounds
is known for long \cite{Jackiw:1976fn,Rubakov:1983bb}.
The mechanism was used to explain the hierarchical mass structure
 of quarks and leptons 
and the suppression of
proton decay \cite{Arkani-Hamed:2000dc,Mirabelli:2000ks,Dvali:2000ha} 
in the context of large extra
dimensions \cite{Arkani-Hamed:1998rs}. 
See also \cite{Grossman:2000ra,Chang:2000nh} 
for fermion localization in the Randall-Sundrum scenario \cite{Randall:1999ee}.
Supersymmetric extension was discussed in \cite{Kaplan:2000av}.

We follow
the formalism of \cite{Arkani-Hamed:2001pv} and 
\cite{Kaplan:2000av} in which four-dimensional  
$N=1$ supersymmetry is 
manifest. Throughout this paper, we consider the five dimensional case where
we have only one extra dimension. We use a unit that the
fundamental five-dimensional Planck scale $M_*=1$.  
Let us denote the coordinate of the  fifth dimension by $y$ and 
introduce five-dimensional fields as one-parameter families of four-dimensional
chiral superfields 
$\Phi(y)$ and its charge conjugated counterpart $\Phi^C(y)$. In components
they are written as $\Phi(y)=\phi(y)+\theta\psi(y)+\theta^2 F(y)$
and $\Phi^C(y)=\phi^C(y)+\theta\psi^C(y)+\theta^2 F^C(y)$, with
$\theta$ being the Grassmann odd coordinate in superspace.
Consider 
the Lagrangian
\begin{equation}
 L=\int dy 
  \left\{ 
      \int d^4 \theta 
         \left( \Phi(y)^{\dagger}\Phi(y) +\Phi^C(y)^{ \dagger}\Phi^C(y)
               \right)
   +  \int d^2 \theta \Phi^C(y) \left[ \partial_y +M(y) \right] \Phi(y)
   +H.c. \right\},
\end{equation}
where $M(y)$ is a field dependent {\em mass term}
\begin{equation}
       M(y) =\Xi(y)+M.
\end{equation}
Here $\Xi$ is another chiral superfield and $M$ is a mass parameter.
In the following we assume that only the scalar component of $\Xi$ has some 
non-vanishing configuration and drop higher components. 
Note that the Lagrangian has only $N=1$ supersymmetry in four dimensions,
while $N=1$ supersymmetry in five dimensions, which corresponds to $N=2$ in
four dimensions, is assumed to be broken, allowing the coupling of $\Xi$
with $\Phi$ and $\Phi^C$.  This assumption is crucial in our purpose.

We make Kaluza-Klein decomposition. 
Equations for zero modes (massless modes in four-dimensional sense) 
are for $\Phi(y)$
\begin{equation}
      \left( \partial_y +M(y) \right) \phi(y)=0, \ \
     \left( \partial_y +M(y) \right) \psi(y)=0,
\end{equation}
and for $\Phi^C(y)$
\begin{equation}
      \left( \partial_y -M(y) \right) \phi^C(y)=0, \ \
     \left( \partial_y -M(y) \right) \psi^C(y)=0.
\end{equation}
Now suppose that the field $\Xi$ has a kink configuration along the
5th dimension. We approximate it as
\begin{equation}
     \Xi(y)= 2 \mu^2 y
\end{equation}
with $\mu^2>0$.
And we assume that only the zero modes for $\phi$ and $\psi$ survive and
those for $\phi^C$ and $\psi^C$ do not.
This may be achieved by appropriate boundary conditions. Then 
the wave functions
are localized around $l\equiv -M/2 \mu^2$ with the following form:
\begin{equation}
  \phi(y)=\psi(y)=
       \left(\frac{2 \mu^2}{\pi} \right)^{1/4}
     \exp\left[-\mu^2 (y-l)^2\right].
\end{equation}

Let us next explain our model and notations. 
It is natural to assume that the five-dimensional Planck scale $M_*$ is
in between the GUT scale $M_{GUT}$ and the four-dimensional Planck scale
$M_{Pl}$. And the compactified 5th dimension is assumed to have length
about one order of magnitude (or so) larger than the inverse of the 
fundamental scale $M_*$. 
To simplify the following argument,  we will not distinguish the two orders of
magnitude difference between $M_{GUT}$ and $M_{Pl}$, and take a unit
$M_{GUT} \approx M_{PL} \approx M_* =1$.

In addition to the Higgs 
multiplets $H(5)$ and $\bar H(5^*)$, we introduce their conjugate multiplets
$H^C(5^*)$ and $\bar H^C(5)$ respectively.  The relevant 
part of the superpotential is 
\begin{eqnarray}
   W&=& \int d^2 \theta \left\{
           H^C(y) \left[ \partial_y +f \Xi(y)+ g \Sigma(y) +M \right] H(y)
        \right.
\nonumber \\
   & & +\left. \bar H^C(y) \left[ \partial_y +\bar{f} 
        \Xi(y)+ \bar{g} \Sigma(y) +\bar{M} \right] \bar{H}(y) \right\},
\end{eqnarray}
where $\Xi$ is a $SU(5)$ singlet, $\Sigma$ is an adjoint Higgs field
responsible for spontaneous breakdown of the $SU(5)$ gauge symmetry,
$f$, $g$, $\bar{f}$, $\bar{g}$ are Yukawa couplings and $M$, $\bar{M}$
are mass parameters.  We assume that  $\Xi$ 
has a non-trivial classical configuration along the fifth dimension. 
To be specific, we suppose in $5\times 5$ matrix notation
\begin{equation}
        \Xi(y)  = 2 \xi^2 (y-y_0) 
          \left( \begin{array}{ccccc}
                1 & & & & \\
                & 1 & & & \\
                & & 1 & & \\
                & & & 1 & \\
                & & & & 1 \end{array} \right).
\label{eq:Xi}
\end{equation}
The effect of $y_0$ in $\Xi(y)$ configuration is absorbed by redefinition
of $M$ and $\bar{M}$ and thus we can take $y_0=0$.
As for $\Sigma(y)$, we assume
\begin{equation}
        \Sigma(y)  = 2 \sigma^2 y
           \left( \begin{array}{ccccc}
                2 & & & & \\
                & 2 & & & \\
                & & 2 & & \\
                & & & -3 & \\
                & & & & -3 \end{array} \right)
             + \Sigma 
              \left( \begin{array}{ccccc}
                2 & & & & \\
                & 2 & & & \\
                & & 2 & & \\
                & & & -3 & \\
                & & & & -3 \end{array} \right).
\label{eq:Sigma}
\end{equation}
Note that the non-trivial dependence on $y$ in the first term of the above
equation is not essential in the subsequent argument. In fact, the
splitting of the triplet and the doublet takes place even for a flat
configuration of $\Sigma(y)$ along the extra dimension, as we will see
shortly.

It follows from Eqs.~(\ref{eq:Xi}) and (\ref{eq:Sigma}) that
\begin{equation}
    f \Xi(y) + g \Sigma(y) +M
  =  \left( \begin{array}{ccccc}
            2 \mu_T^2 y+M_T & & & & \\
            & 2 \mu_T^2 y+M_T & & & \\
            & & 2 \mu_T^2 y+M_T & & \\
            & & & 2 \mu_{u}^2 y+M_u & \\
            & & & & 2 \mu_u^2 y+M_u \end{array} \right),
\end{equation}
where
\begin{eqnarray}
   \mu_T^2 & \equiv & f \xi^2 + 2 g \sigma^2,
\\
   \mu_u^2 & \equiv & f \xi^2 - 3 g \sigma^2,
\\
   M_T & \equiv & M+ 2 g \Sigma,
\\
   M_u & \equiv & M- 3 g \Sigma.
\end{eqnarray}
A similar expression is obtained for $\bar H$ with
\begin{eqnarray}
     \mu_{\bar T}^2 & \equiv & \bar f \xi^2 + 2 \bar g \sigma^2,
\\
   \mu_d^2 & \equiv & \bar f \xi^2 - 3 \bar g \sigma^2,
\\
    M_{\bar T} & \equiv & \bar{M} +2 \bar{g} \Sigma,
\\
    M_d & \equiv & \bar M - 3 \bar g \Sigma.         
\end{eqnarray}

 After the $SU(5)$ breakdown,
the 5 dimensional Higgs $H$ is decomposed into the color triplet $H_T$ and
the doublet Higgs $H_u$. In our setting the zero mode wave functions
for 
$H_T$ and $H_u$
become
\begin{eqnarray}
  \psi_T(y) &=& \left( \frac{2 \mu_T^2}{\pi} \right)^{1/4}
              \exp\left[ -\mu_T^2 \left( y-l_T \right)^2 \right],
\\
  \psi_u(y) &=& \left( \frac{2 \mu_u^2}{\pi} \right)^{1/4}
              \exp\left[ -\mu_u^2 \left( y-l_{u} \right)^2 \right],
\end{eqnarray}
where $l_T=-M_T/2\mu_T^2$ and $l_u=-M_u/2\mu_u^2$.  Similarly we obtain the
wave functions for $H_{\bar T}$ and 
$H_d$, by replacing $\mu_T$ and $\mu_u$ with
$\mu_{\bar T}$ and $\mu_d$. Explicitly they are
\begin{eqnarray}
  \psi_{\bar T}(y) &=& \left( \frac{2 \mu_{\bar T}^2}{\pi} \right)^{1/4}
              \exp\left[ -\mu_{\bar T}^2 \left( y-l_{\bar T} \right)^2 \right],
\\
  \psi_d(y) &=& \left( \frac{2 \mu_d^2}{\pi} \right)^{1/4}
              \exp\left[ -\mu_d^2 \left( y-l_{d} \right)^2 \right]
\end{eqnarray}
with $l_{\bar T}=-M_{\bar T}/2\mu_{\bar T}^2$ and $l_d=-M_{d}/2\mu_d^2$.

Let us next consider the triplet-doublet splitting in this setting.
What we would like to realize 
is a situation that the triplet is superheavy while
the doublet is light.  To achieve this, we consider the following two cases. 

\begin{figure}[ht]
  \begin{center}
    \makebox[0cm]{
      \scalebox{1.0}{
        \includegraphics{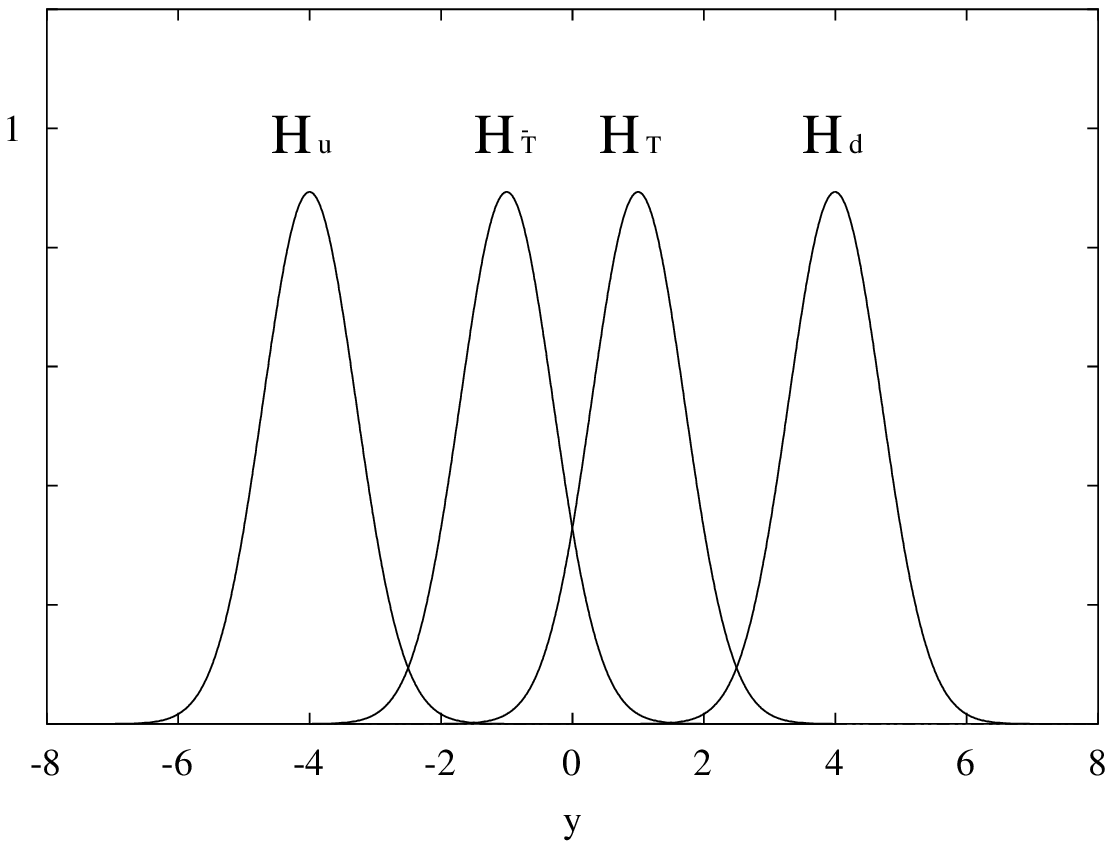}
        }
      }
    \caption{Higgs profiles for 
      $f = \bar{f} = g = - \bar{g} = 1, \sigma = 0, \xi = M' = M_*$
      and,  $M = - \bar{M} = - \Sigma = 2 M_*$,
      so that triplet Higgs masses $\sim 0.14 M_*$ 
      and doublet ones $\sim 1.3 \times 10^{-14} M_*$.}
    \label{fig:std1}
  \end{center}
\end{figure}

The first case is that $H$ and $\bar H$ have a five dimensional mass term
and overlaps of the wave functions themselves determine the size of the 
masses in four dimensions. The five dimensional mass term is given by
the following superpotential
\begin{equation}
   \int d^2 \theta \int dy  M' H(y) \bar H(y),
\label{eq:mass-term}
\end{equation}
where $M'$ is the mass parameter of the order of $M_*$.
In this case the smallness of the doublet mass should be explained by the
small overlap of the two wave functions $\psi_u(y)$ and $\psi_d(y)$. The
pattern of the wave functions we wish to get is illustrated in 
\Figref{fig:std1}. 
Requiring the resulting mass term does not exceed the electroweak scale,
we find
\begin{equation}
   \int dy \psi_u(y) \psi_d(y) \lesssim 10^{-(14 \sim 16)}.
\end{equation}
The right hand side of the above equation is computed as
\begin{equation}
    \int dy \psi_u(y) \psi_d(y)=
  \left( \frac{4\mu_u^2 \mu_d^2}{(\mu_u^2+\mu_d^2)^2} \right)^{1/4}
        \exp\left[ -\frac{\mu_u^2 \mu_d^2}{(\mu_u^2+\mu_d^2)}
                    (l_u -l_d)^2 \right],
\end{equation}
from which we obtain the constraint
\begin{equation}
   | l_u - l_d| \gtrsim 6 
       \left( \frac{(\mu_u^2+\mu_d^2)}{\mu_u^2 \mu_d^2} \right)^{1/2}.
\end{equation}
It is possible to arrange the locations of the wave functions to
satisfy the above constraint. It only requires a tuning of the
parameters of order at most 10.  A potential problem of this configuration 
where the triplets are located in between the two doublets is
that the triplet Higgs substantially couples to the quarks and leptons when
they are located to yield realistic Yukawa couplings and thus a mechanism
to suppress the dimension five proton decay given below  will not work. 

\begin{figure}[ht]
  \begin{center}
    \makebox[0cm]{
      \scalebox{1.0}{
        \includegraphics{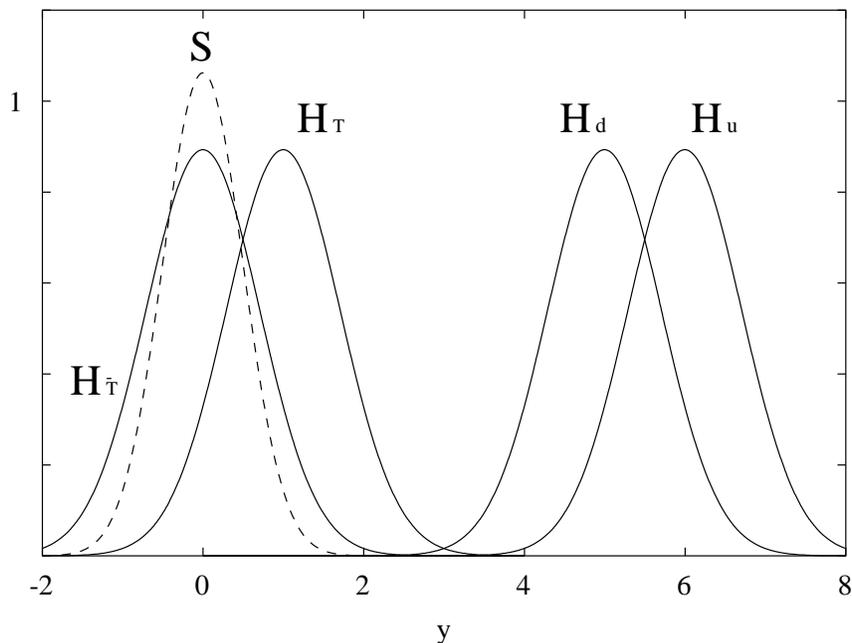}
        }
      }
    \caption{Higgs and singlet field profiles for 
      $f = \bar{f} = g = \bar{g} = 1, \sigma = 0, 
      \xi = M_*, \Sigma = 2 M_*$,
      $M = -6 M_*, \bar{M} = -4 M_*, \mu_s = \sqrt{2} M_*$ and, $l_s = 0$,
      so that $\lambda_2 \sim 3.3 \times 10^{-14}$ 
      and $\lambda_3 \sim 0.35$. A singlet $S$ which is localized around
      the locations of the triplets is introduced.}
   \label{fig:std2}
  \end{center}
\end{figure}

The second possibility we want to discuss here is to introduce a new
singlet field $S(y)$. Suppose that, instead of the direct mass term considered 
above, the singlet has the following Yukawa interaction with $H$ and $\bar H$
\begin{equation}
    \lambda \int d^2 \theta \int dy S(y) H(y) \bar H(y).
\label{eq:singlet-coupling}
\end{equation}
The presence of Eq.~(\ref{eq:singlet-coupling}) and the absence of 
Eq.~(\ref{eq:mass-term}) may be a consequence of some symmetry. 
Assume that the $S$ field is localized nearby  $H_T$ and $H_{\bar T}$ 
peaks while
the peaks of $H_u$ and $H_d$ are located far away.  The configuration
is schematically depicted in \Figref{fig:std2}.
The realization requires again some
tuning of the parameters of order at most 10. This should be compared to the
huge fine tuning of $10^{14}$ or so which is needed in the 
conventional approach to the minimal 
SUSY $SU(5)$.
Separation of $S$ and $H_u$, $H_d$ will imply small Yukawa coupling for them 
in four dimensions. Thus the dimensional reduction yields the following
effective theory
\begin{equation}
   \int d^2 \theta \left[ 
  \lambda_3 S H_T H_{\bar T} + \lambda_2 S H_u H_d \right],
\end{equation}
where $\lambda_2$ and $\lambda_3$ are Yukawa couplings which we assume to 
achieve $\lambda_2 \lesssim 10^{-14}$ and 
$\lambda_3\sim O(1) $. 
Note that in the resulting four-dimensional theory  the $SU(5)$ 
symmetry is not respected. Recall that This is due to the fact that the 
the $SU(5)$ breaking Higgs field nontrivial couplings to the Higgs multiplets 
in five dimensions, which makes the wave functions of the triplet and the
doublet completely different.

On the other hand, the (unbroken) gauge fields in the standard model have
flat configuration in the extra dimension and hence the gauge coupling
unification is guaranteed at tree level. It is interesting
to see how Kaluza-Klein modes affect the coupling unification, which is beyond
the scope of this paper.

If $S$ condensates around or just below the GUT scale, then $H_T$, 
$H_{\bar T}$ 
acquires a GUT scale mass, while the doublet remains very light. Thus
the triplet-doublet splitting is achieved. 
The singlet $S$ will become superheavy and thus it does not destabilize
the gauge hierarchy at quantum level \cite{light-singlet}.
Note that the splitting
in our setting is not automatic, but requires some adjustment of the
parameters to fix the locations of the wave functions. In this sense,
our mechanism  is similar to
the conventional case of the minimal SUSY $SU(5)$. However, the amount
of the adjustment required here is not very huge, thanks to the Gaussian
behaviour of the wave functions.

Realistic Yukawa couplings in the quark sector can be
obtained by using the localization of the chiral multiplets. 
Hierarchically small Yukawa couplings are explained by small overlaps of 
the wave functions.  
An example
is given in \cite{Kaplan:2000av} in SUSY $SU(5)$. Another example in which
a variant of Fritzsch-type texture \cite{Fritzsch:1978vd} 
is obtained  will be discussed elsewhere 
\cite{KakizakiYamaguchi}. 

Another interesting point we should stress is about proton decay. In
SUSY GUTs the dimension five operators which cause proton decay can be
induced  by exchange of the triplet Higgs multiplets. As was discussed
earlier, this is one of the major obstacles to build a realistic SUSY
GUT model. In most cases, the Yukawa couplings of the triplet Higgs
are similar in size to those of the doublet partner.  In our setting,
however, this may not be the case 
because the $SU(5)$ invariance is not manifest in the
low-energy four-dimensional theory. It provides a very intriguing
resolution of the proton decay problem. In fact,
the Yukawa couplings between the triplet Higgs and
quarks/leptons can be suppressed if the distance between the locations
of quarks/leptons and the location of the triplet Higgs is so large
that the convolution of the wave functions becomes sufficiently small.
Then the proton decay by the triplet Higgs exchange can be suppressed
to a negligible level.

To summarize, we have proposed a new approach to the triplet-doublet splitting
in SUSY GUT based on higher dimensional theories. If the GUT symmetry breaking
is inherently higher dimensional, the four-dimensional effective theory obtained from dimensional reduction need not have GUT symmetry phase.   Then the mass
of the doublet Higgs can be completely different from that of the triplet
counterpart.  We have used the mechanism of field localization in the extra
dimension under kink configuration to realize this idea in field theoretic
approach.  The GUT braking Higgs is assumed to have non-vanishing couplings
to the 5 and $5^*$ Higgs multiplets and thus plays a non-trivial role in the localization process.  This  splits
the wave functions of the doublet Higgses from those of the triplet.
To obtain very light doublet and superheavy triplet, we have considered the 
two cases.  The latter case with introduction of a new singlet is particularly 
interesting, since it can provide realistic quark masses by appropriate
localization of quark wave functions. Furthermore we can explain 
the suppression of the proton decay through colored-triplet exchange 
by making the Yukawa
couplings of the triplet to quarks and leptons very small.  
This can be achieved
if the triplet Higgs fields are localized far away from the positions of
the quarks and leptons.  This is another advantage of our mechanism that the
four-dimensional theory does not have the GUT symmetric phase.

\section*{Acknowledgment} 
This work was supported in part by the
Grant-in-aid from the Ministry of Education, Culture, Sports, Science
and Technology, Japan, priority area (\#707) ``Supersymmetry and
unified theory of elementary particles,'' and in part by the
Grants-in-aid No.11640246 and No.12047201.


%
%
%
\end{document}